\documentclass[prl,aps,twocolumn,showpacs,superscriptaddress]{revtex4-1}
\usepackage{hyperref}
\hypersetup{colorlinks=true, citecolor=blue, filecolor= black, linkcolor= blue, urlcolor= blue}
\usepackage{graphicx}
\usepackage{dcolumn}
\usepackage{bm}
\begin{document}


\title{Semimetallic Two-Dimensional Boron Allotrope with Massless Dirac Fermions}

\author{Xiang-Feng Zhou}
\email{xfzhou@nankai.edu.cn}
\email{zxf888@163.com}
\affiliation{School of Physics and Key Laboratory of Weak-Light Nonlinear Photonics, Nankai University, Tianjin 300071, China}
\affiliation{Department of Geosciences, Center for Materials by Design, and Institute for Advanced Computational Science, Stony Brook University, Stony Brook, New York 11794, USA}
\author{Xiao Dong}
\affiliation{School of Physics and Key Laboratory of Weak-Light Nonlinear Photonics, Nankai University, Tianjin 300071, China}
\affiliation{Department of Geosciences, Center for Materials by Design, and Institute for Advanced Computational Science, Stony Brook University, Stony Brook, New York 11794, USA}

\author{Artem R. Oganov}
\affiliation{Department of Geosciences, Center for Materials by Design, and Institute for Advanced Computational Science, Stony Brook University, Stony Brook, New York 11794, USA}
\affiliation{Moscow Institute of Physics and Technology, 9 Institutskiy lane, Dolgoprudny city, Moscow Region 141700, Russian Federation}
\affiliation{School of Materials Science, Northwestern Polytechnical University, Xi'an 710072, China}

\author{Qiang Zhu}
\affiliation{Department of Geosciences, Center for Materials by Design, and Institute for Advanced Computational Science, Stony Brook University, Stony Brook, New York 11794, USA}

\author{Yongjun Tian}
\affiliation{State Key Laboratory of Metastable Materials Science and Technology, Yanshan University, Qinhuangdao 066004, China}

\author{Hui-Tian Wang}
\affiliation{School of Physics and Key Laboratory of Weak-Light Nonlinear Photonics, Nankai University, Tianjin 300071, China}
\affiliation{National Laboratory of Solid State Microstructures, Nanjing University, Nanjing 210093, China}

\begin{abstract}
\noindent It has been widely accepted that planar boron structures, composed of triangular and hexagonal motifs are the most stable two dimensional (2D) phases and likely precursors for boron nanostructures. Here we predict, based on \textit{ab initio} evolutionary structure search, novel 2D boron structure with non-zero thickness, which is considerably, by 50 meV/atom lower in energy than the recently proposed $\alpha$-sheet structure and its analogues. In particular, this phase is identified for the first time to have a distorted Dirac cone, after graphene and silicene the third elemental material with massless Dirac fermions. The buckling and coupling between the two sublattices not only enhance the energetic stability, but also are the key factors for the emergence of the distorted Dirac cone.
\end{abstract}



\maketitle
Boron is a fascinating element because of its chemical and structural complexity. There are now at least 5 known polymorphs (structural forms), yet the ground-state structure of boron until recently was controversial \cite{R01}. While at least 16 forms of boron (some of which were probably impurity-stabilized) have been reported, the existence as pure polymorphs of boron has been established for the $\alpha$ rhombohedral, $\beta$ rhombohedral, two tetragonal phases, and the recently discovered orthorhombic high-pressure partially ionic $\gamma$ phase \cite{R02}. Boron has been investigated both theoretically and experimentally as bulk boron, nanotubes, clusters, quasi planar, monolayer, and bilayer sheets \cite{R02,R03,R04,R05,R06,R07,R08,R09,R10,R11,R12,R13,R14,R15,R16,R17,R18,R19,R20}. Novel boron nanobelts or nanowires have been successfully synthesized \cite{R13,R14}, and experimentally shown to be semimetals or narrow-gap semiconductors, but the exact atomic structures are still not fully resolved \cite{R14}. So far, planar geometry was not seen in boron crystals, which are built instead of B$_{12}$ icosahedra \cite{R06}. Boron sheets with buckled triangular arrangement are thought to be most favorable, and consequently used to construct boron nanotubes \cite{R18}. Recently, a new class of boron sheets composed of triangular and hexagonal motifs, exemplified by the so called $\alpha$-sheet structure, have been identified to be energetically most stable \cite{R06}. This has also successfully explained the proposed stability of B$_{80}$ fullerenes \cite{R05}. Since the boron sheets can serve as a building block (or precursor) for fullerenes, nanotubes, and nanoribbons, understanding its structure and stability is a prerequisite for all those nanostructures \cite{R12}. Using particle swarm optimization technique, some planar similar structures were predicted to have the same or slightly lower energy than the $\alpha$-sheet \cite{R19,R20}. However, recently the stability of B$_{80}$ fullerene was challenged \cite{R11}. Furthermore, the $\alpha$-sheet is dynamically unstable and transforms to its analogues (non-planar $\alpha'$-sheet) by removing the soft mode near the M(0.5 0 0) point \cite{R20}. Such questions spur us to explore other potentially stable structures or structures with novel electronic properties by first-principles calculations.

Structure searches were performed using the \textit{ab initio} evolutionary algorithm \textsc{uspex} \cite{R21,R22,R23} which has been successfully applied to various bulk materials \cite{R24,R25,R26}. The extension to 2D structure prediction has been implemented and is now available in the \textsc{uspex} code. In these calculations, initial structures are randomly produced using plane group symmetry with a user-defined initial thickness (the energetic stability is sensitive to the constraint of thickness; we want to study the monolayer and bilayer 2D structures mostly, according to the experimental evidence for the spacing ($\sim$ 3.2 {\AA}) between two adjacent layers of multiwalled boron nanotubes \cite{R15}, the initial thickness was set to 3 {\AA}, and allowed to change during relaxation), and all newly produced structures are relaxed, and relaxed energies are used for selecting structures as parents for the new generation of structures (produced by well-designed variation operators, such as heredity and softmutation). The target is to find the most stable 2D structures. The structural relaxations use the all electron projector augmented wave method \cite{R27} as implemented in the Vienna \textit{ab initio} simulation package (VASP) \cite{R28}. The exchange-correlation energy was treated within the generalized gradient approximation (GGA), with the functional of Perdew, Burke, and Ernzerhof (PBE) \cite{R29}. In addition, the hybrid HSE06 functional with the screening parameter ($\omega$) of 0.2 {\AA}$^{-1}$ is also employed to confirm the energetic stability and the band structures of several 2D boron structures \cite{R30}. A cutoff energy of 450 eV and a Monkhorst-Pack Brillouin zone sampling grid with resolution of $2 \pi \times 0.04$ \AA$^{-1}$ is used. Phonon dispersion curves were computed with the \textsc{quantum-espresso} package \cite{R31} using the PBE functional, ultrasoft potential, a cutoff energy of 50 Ry for the wave functions, a $5 \times 5 \times 1$ \ $q$-point mesh for the $\alpha$-sheet, and $4 \times 6 \times 1$ \ $q$-point meshes for other structures.

\begin{figure}[h]
\begin{center}
\includegraphics[width=8cm]{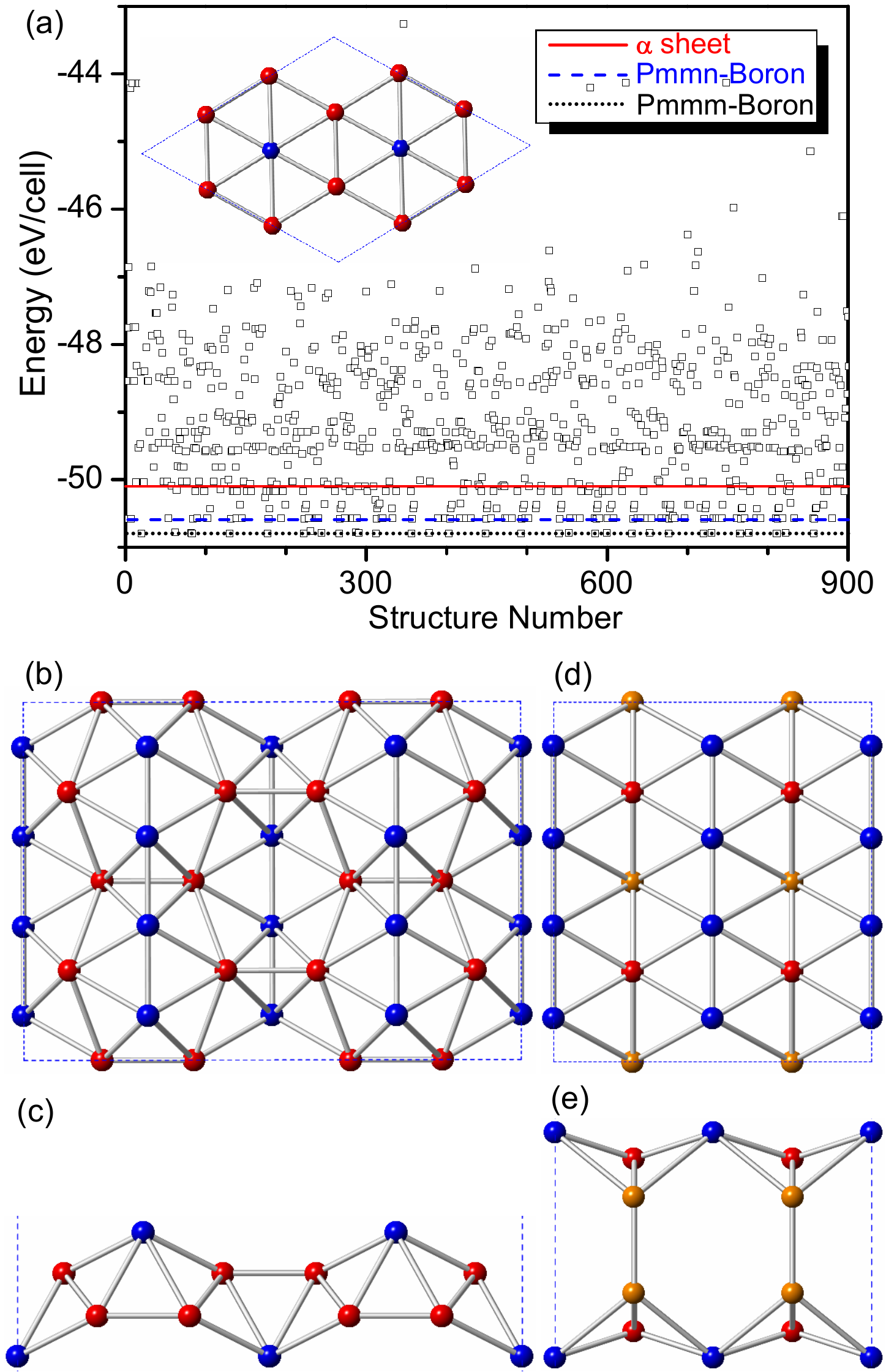}
\caption{%
(Color online) (a) Typical enthalpy evolution for an 8-atom 2D boron system during an evolutionary structure search. The inset shows the structure of the $\alpha$-sheet. (b) and (c) projections of $2 \times 2 \times 1$ supercell of $Pmmn$-boron structure along [001] and [100] directions. (d) and (e) projections of $2 \times 2 \times 1$ supercell of $Pmmm$-boron structure along [001] and [100] directions. The nonequivalent atomic positions for boron polymorphs are shown by different colors.}
\end{center}
\end{figure}

\begin{table}
\caption{Calculated lattice constants, atomic positions, and the total energy ($E_t$) of the polymorphs of boron from GGA (PBE) results; the $\alpha$-sheet and $\alpha$-boron have $P6/mmm$ and $R\bar{3}m$ symmetry. The experimental values for $\alpha$-boron (from Ref. [33]) are also listed for comparison.}
\begin{tabular}{lllllc}
\hline\hline
Phase & $a$    & $b$    & $c$    & Atomic positions & $E_t$ \\
      &({\AA}) &({\AA}) &({\AA}) &                  & (eV/atom) \\
\hline
$\alpha$-sheet & 5.07 & 5.07 & 13.00 & B1 (0.667 0.333 0.5)   & -6.28 \\
               &      &      &       & B2 (0.667 0.667 0.5)   &  \\
$Pmmn$         & 4.52 & 3.26 & 13.00 & B1 (0.5 0.753 0.584)   & -6.33 \\
               &      &      &       & B2 (0.185 0.5 0.531)   &  \\
$Pmmm$         & 2.88 & 3.26 & 13.00 & B1 (0.0 0.243 0.657)   & -6.36 \\
               &      &      &       & B2 (0.5 0.5 0.621)     &  \\
               &      &      &       & B3 (0.5 0.0 0.567)     &  \\
$\alpha$-boron & 4.90 & 4.90 & 12.55 & B1 (0.803 0.197 0.976) & -6.68 \\
               &      &      &       & B2 (0.119 0.238 0.892) &  \\
Experiment     & 4.91 & 4.91 & 12.57 & B1 (0.804 0.197 0.976) &  \\
               &      &      &       & B2 (0.118 0.235 0.893) &  \\
\hline\hline
\end{tabular}
\end{table}

The searches were performed with 6, 8, 10, 12, 14, 16, and 18 atoms per unit cell. We find that two special structures (designated as 2D-B$_{14}$ and 2D-B$_{16}$, which contain 14 and 16 atoms per unit cell, respectively) are much lower in energy than the $\alpha$-sheet \cite{R32}. The true thickness (the distance between the two planes which include the highest and lowest atomic positions) of 4.544 {\AA} and 6.361 {\AA} for the 2D-B$_{14}$ and 2D-B$_{16}$ phases is responsible for their superior energetic stability. In general, 2D structures become more stable with increasing their thickness as they approach the bulk state. There are too many low-energy structures found by \textsc{uspex}, some possess no special electronic structure, and will not be discussed further. We focus on the low-energy monolayer and bilayer structures, which have considerable chances of being prepared experimentally on a suitably chosen substrate. In this letter, we present a typical example (8 atoms/cell) shown in Fig. 1. It will be a benchmark case for the 8-atom system because the $\alpha$-sheet also contains 8 atoms per unit cell. This allows one to test whether the $\alpha$-sheet or a better 8-atom structure is found in the search. Also, it is important to know whether there are metastable structures with novel electronic properties. Indeed, the $\alpha$-sheet structure is reproduced during the search, as shown in Fig. 1(a). Among the structures that are more stable than the $\alpha$-sheet, there are many low-symmetry (e.g., $P\bar{1}$) structures, indicating that 2D-boron is a frustrated system. From Fig. 1(a), the two most stable symmetric phases are designated as $Pmmn$-boron and $Pmmm$-boron. Table I lists the lattice constants, atomic positions, and total energies of $Pmmn$-boron, $Pmmm$-boron, $\alpha$-sheet, and $\alpha$-boron. The calculated ground state lattice constant of the bulk $\alpha$-boron is in excellent agreement with the experimental value \cite{R33}, which establishes the reliability and accuracy of the GGA-PBE calculations. The GGA-PBE results show that $Pmmm$-boron and $Pmmn$-boron are 0.08 eV/atom and 0.05 eV/atom lower in energy than $\alpha$-sheet structure, but are 0.32 eV/atom and 0.35 eV/atom higher in energy than bulk $\alpha$-boron, indicating that the two 2D phases are (as expected) metastable. Moreover, the HSE06 calculations show the total energies for $\alpha$-sheet, $Pmmn$-boron, $Pmmm$-boron, and bulk $\alpha$-boron are -6.94, -7.03, -7.05, and -7.43 eV/atom compared with the corresponding values of -6.28, -6.33, -6.36, and -6.68 eV/atom from GGA-PBE calculations, i.e, GGA-PBE and HSE06 give the same ranking of structures by stability. The structure of $Pmmn$-boron has two nonequivalent atomic positions (or two sublattices), as illustrated in Fig. 1(b) and 1(c) by different colors. The most stable $Pmmm$-boron is made of buckled triangular layers [Fig. 1(d) and 1(e)], and has three nonequivalent atomic positions. Compared with the planar $\alpha$-sheet, there are common characters (buckling and coupling) for the geometric structure of both $Pmmn$-boron and $Pmmm$-boron.

An Aufbau principle was proposed whereby the most stable structures should be composed of buckled triangular motifs \cite{R18}. Experiments on small clusters of 10--15 atoms support this view \cite{R03}. From our prediction of 2D boron phases, all the most stable structures are also made of buckled triangular layers. The buckling is formed within a given thickness. It can mix in-plane and out-of-plane states and can be thought of as a symmetry-reducing distortion that enhances binding by opening a band gap or pseudogap \cite{R06}. 2D-boron is a frustrated system, which tends to have many complex near-ground-state structures, and such systems violate the correlation between the energy and geometric simplicity of crystal structures (simpler structures are statistically more stable) \cite{R34}. This is also one of the most important reasons to explain the dynamical instability of planar $\alpha$-sheet. The coupling between different sublattices leads to the formation of strong covalent B-B bonds, enhancing structural stability. Therefore, non-zero thickness is responsible for the energetic stability of the 2D boron polymorphs. Similarly, in a recent paper, it was found that BH, a novel high-pressure phase, adopts semiconducting structures with buckled boron layers broadly similar to the ones presented here and passivated by hydrogen atoms and only at ultrahigh pressures above 168 GPa transforms into a metallic phase with flat triangular boron layers \cite{R26}.

\begin{figure}[h]
\begin{center}
\includegraphics[width=8cm]{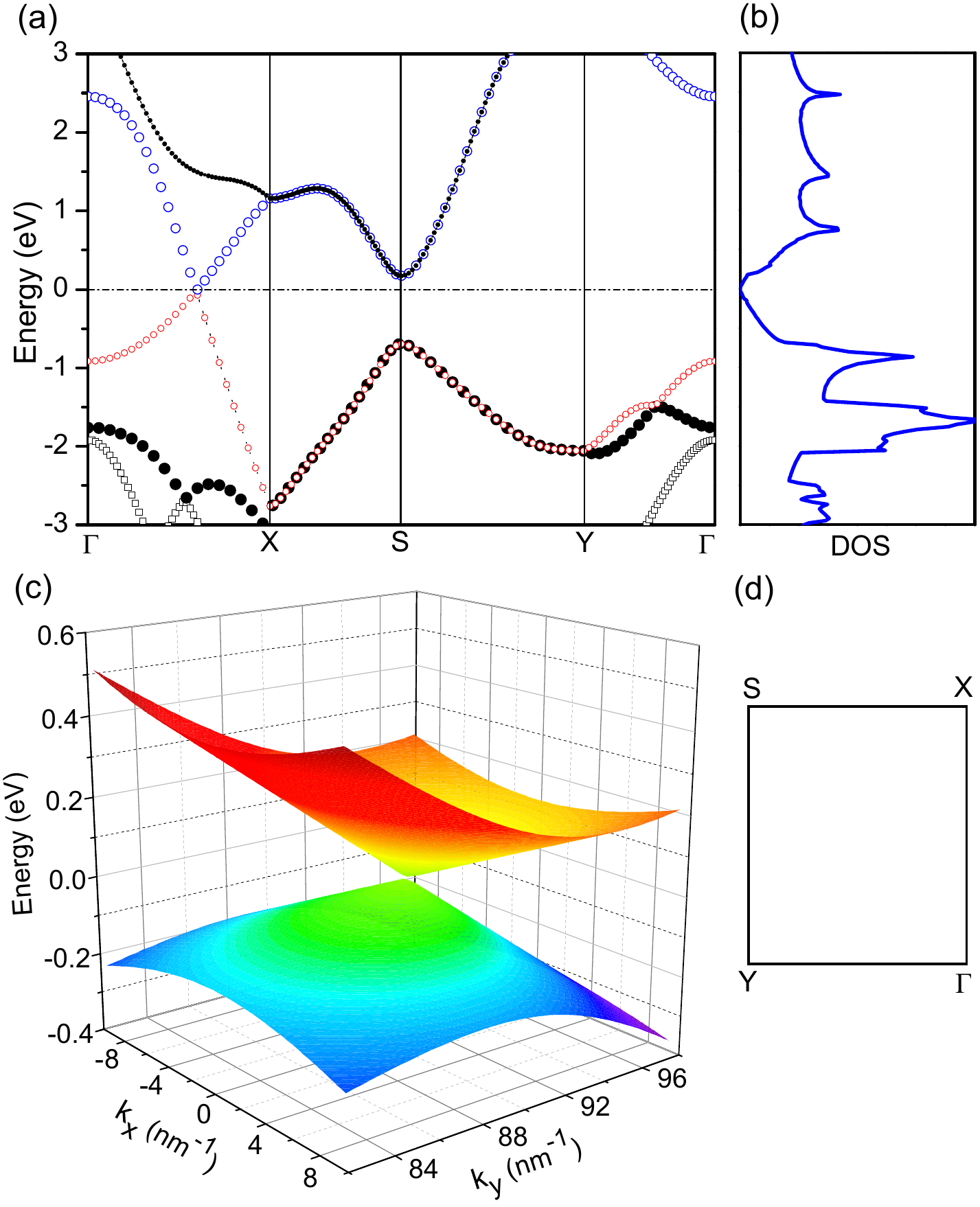}
\caption{%
(Color online) Electronic structure of $Pmmn$-boron. (a) Band structure, (b) DOS, (c) Dirac cone formed by the valence and conduction band in the vicinity of the Dirac point, (d) First Brillouin zone with the special $k$ points: $\Gamma$(0 0 0), X(0 0.5 0), S(-0.5 0.5 0), and Y(-0.5 0 0).}
\end{center}
\end{figure}

Both the $\alpha$-sheet and the $Pmmm$-boron are metallic \cite{R06,R32}. By comparison, the $Pmmn$-boron is a zero-gap semiconductor. Its band structure [see Fig. 2(a)] shows valence and conduction bands meeting in a single point (0 0.3 0) at the Fermi level. The density of states (DOS) of $Pmmn$-boron is zero at the Fermi level. So this meeting point is a Dirac point which is elaborated in Fig. 2(c). The valence and the conduction band of $Pmmn$-boron in the vicinity of the Dirac point show the presence of a distorted Dirac cone which is very similar to that of 6, 6, 12-graphyne \cite{R35}. These bands exhibit a linear dispersion in both $k_x$ and $k_y$ directions, i.e.\ like in graphene, the effective mass of the mobile electron is zero.  The slope of the bands in the $k_x$ direction is $\pm$23 eV {\AA}, equivalent to a Fermi velocity $\partial E$/$\partial k_x$ = $0.56 \times 10^6$ m/s. In the $k_y$ direction, the slope of the bands equal --48 eV {\AA} ($v_{Fy}$ = $1.16 \times 10^6$ m/s) and 19 eV {\AA} ($v_{Fy}$ = $0.46 \times 10^6$ m/s), compared to $\pm$34 eV {\AA} ($v_F$ = $0.82 \times 10^6$ m/s) in graphene when approaching a Dirac point along the $\Gamma$ -- $K$ line. \cite{R35}. The anisotropy of the Dirac cones with different slopes at the Dirac points in the $k_x$ and $k_y$ directions, implies direction-dependent electronic properties. Figure 2(d) shows the first Brillouin zone of $Pmmn$-boron with special $k$-points, and also indicates that hexagonal symmetry is not a prerequisite for the existence of a Dirac cone: here we deal with an orthorhombic ($Pmmn$) structure that has two inequivalent boron sites. We also should note that the distorted Dirac cone of $Pmmn$-boron is a robust feature and is found also when using the HSE06 functional, its position shifts 0.18 eV above the Fermi level \cite{R32}.

\begin{figure}[h]
\begin{center}
\includegraphics[width=8cm]{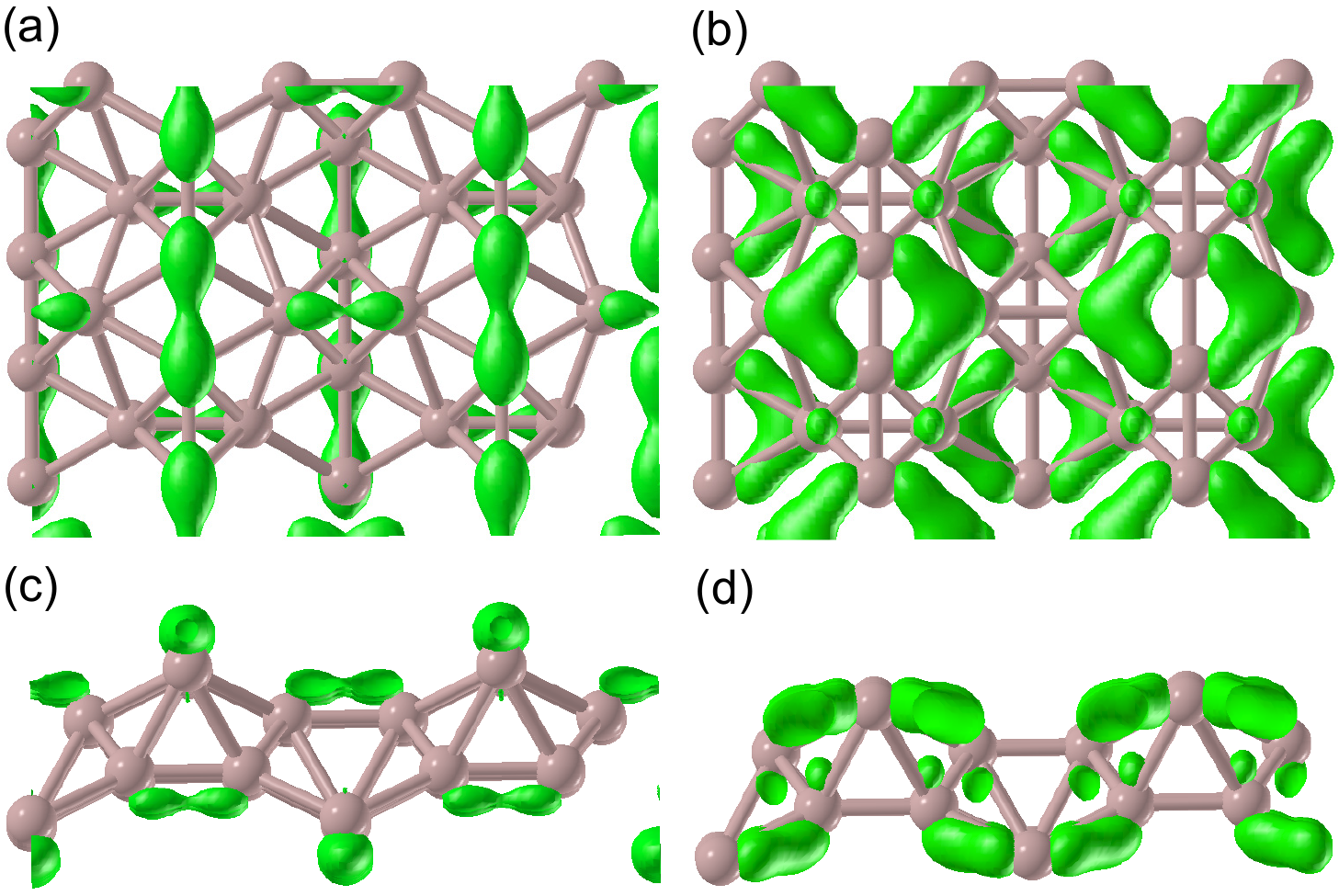}
\caption{%
(Color online) The band decomposed charge density of $Pmmn$-boron at the Dirac point: (a) and (c) projections of the charge density of the highest valence band along [001] and [100] directions; (b) and (d) projections of the charge density of the lowest conduction band along [001] and [100] directions.}
\end{center}
\end{figure}

To explore the physical origin of the Dirac cone, the band-decomposed charge density at Dirac point is plotted in Fig. 3. Figures 3(a) and 3(c) show the charge density of the highest valence band at the Dirac point along [001] and [100] directions. The charge density distribution is derived from the out-of-plane ($p_z$ orbitals) states of two sublattices. The hybrids of in-plane states and out-of-plane states between two sublattices are responsible for the charge density distribution of the lowest conduction band at the Dirac point, see Fig. 3(b) and 3(d). The charge density distributions for both the conduction band and the valence band at the Dirac point have mirror symmetry along $x$ and $y$ directions. The hybrids of in-plane ($p_x$ orbitals from the buckled boron chains) and out-of-plane states ($p_z$ orbitals from the buckled irregular boron hexagons) are a unique feature responsible for the emergence of Dirac cone. In addition, for $Pmmn$-boron, there are 2 kinds of B-B bonds between two sublattices (buckled chains and hexagons) with bond lengths of 1.80{\AA} and 1.89 {\AA}. The hybrids mostly take place in the short B-B bonds (1.80 {\AA}) between two sublattices. The origin of Dirac cone of $Pmmn$-boron is different from those of graphene \cite{R36,R37}, $T$ graphene \cite{R38}, and graphynes \cite{R35} where they arise from the crossing $\pi$ and $\pi^{\ast}$ bands derived from $p_z$ orbitals exclusively.

The phonon dispersion curves and phonon density of states (PDOS) show that $Pmmm$-boron and $Pmmn$-boron are dynamically stable (see Fig. 4). The structure of $Pmmn$-boron contains triangular B$_3$ units that condense into fragments of B$_{12}$ icosahedra, so ubiquitously found in all known boron allotropes (none of which have layered structures). The motif of $Pmmn$-boron can be thought of overlapping buckled pentagonal and hexagonal pyramids, which are also very close to the structural fragments of bulk $\alpha$-boron. Bader charges show that the charge transfer between two sublattices of $Pmmn$-boron is $\pm$0.05 $e$ \cite{R39}, similar to that of $\alpha$-boron ($\pm$0.056 $e$) \cite{R02}. Moreover, the lattice constants of $Pmmn$-boron match very well with the (110) plane of some metals or metal oxides, it may be expected to be synthesized by depositing boron atoms on certain metal substrates, which has been applied in the preparation of graphene \cite{R40}. All of these suggest that $Pmmn$-boron or its derivatives may be made in nanostructures or thin films. Interestingly, some boron nanobelts of unknown atomic-scale structure show a band gap of 0.2 $\pm$0.2 eV \cite{R14}. $Pmmn$-boron shows that the existence of a Dirac cone is not a unique feature of carbon based materials, such as graphene.

\begin{figure}[h]
\begin{center}
\includegraphics[width=8cm]{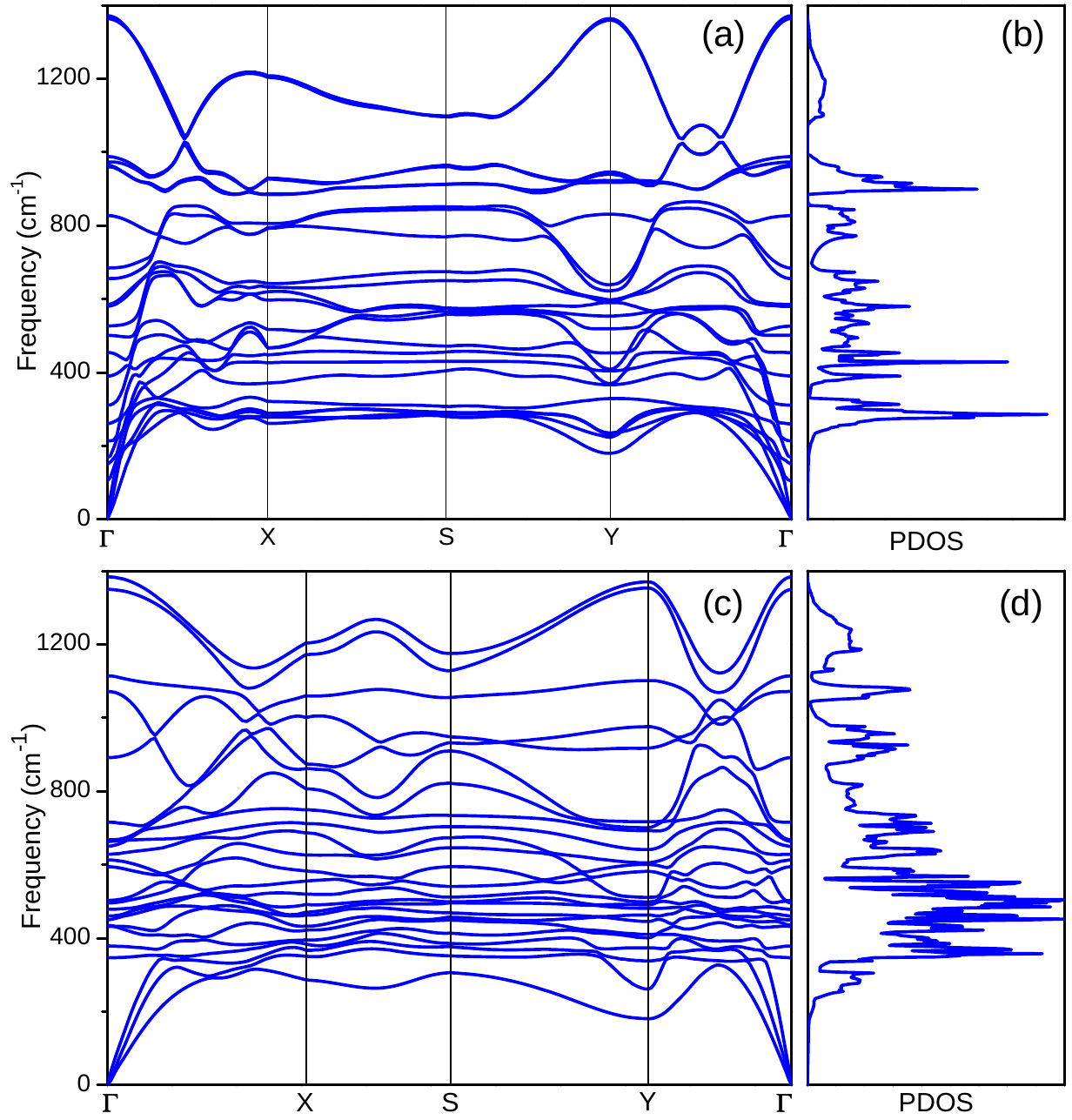}
\caption{%
(Color online) [(a) and (b)] Phonon dispersion and PDOS of $Pmmm$-boron at ambient pressure. [(c) and (d)]Phonon dispersion and PDOS of $Pmmn$-boron at ambient pressure.}
\end{center}
\end{figure}

In summary, a systematic structure search for 2D phases of boron identified two orthorhombic structures with space groups $Pmmm$ and $Pmmn$, which may be kinetically stable at ambient conditions. The $Pmmn$ structure is lower in energy than the earlier reported planar structures due to non-zero thickness. Most strikingly, this structure is identified to have a distorted Dirac cone, the first in non-graphenelike 2D materials. The quasiparticle group velocity is comparable to that in graphene, but strongly direction-dependent. The buckling and coupling between the two constituent sublattices are the key factors for the energetic stability and the emergence of the distorted Dirac cone. Our findings suggest that the current design strategy for boron sheets and nanotubes has to include a finite thickness of boron layers.

X.F.Z. thanks Philip B. Allen for reading the manuscript before submission, and Pengcheng Chen for valuable discussions. This work was supported by the National Science Foundation of China (Grants No. 11174152, No. 51332005, and No. 91222111), the National 973 Program of China (Grant No. 2012CB921900), the Program for New Century Excellent Talents in University (Grant No. NCET-12-0278), and the Fundamental Research Funds for the Central Universities (Grant No. 65121009). A.R.O. thanks the National Science Foundation (EAR-1114313, DMR-1231586), DARPA (Grants No. W31P4Q1310005 and No. W31P4Q1210008), DOE (Computational Materials and Chemical Sciences Network (CMCSN) project DE-AC02-98CH10886), CRDF Global (UKE2-7034-KV-11), AFOSR (FA9550-13-C-0037), the Government (No. 14.A12.31.0003) and the Ministry of Education and Science of Russian Federation (Project No. 8512) for financial support, and Foreign Talents Introduction and Academic Exchange Program (No. B08040). Calculations were performed on XSEDE facilities and on the cluster of the Center for Functional Nanomaterials, Brookhaven National Laboratory, which is supported by the DOE-BES under contract no. DE-AC02-98CH10086.



\end{document}